\documentclass[twocolumn,pre,showpacs]{revtex4}
\usepackage{amsmath}

\begin{document}

\title{Sound damping in ferrofluids:\\
Magnetically enhanced compressional viscosity }
\author{ Hanns Walter M\"{u}ller}\email{hwm@mpip-mainz.mpg.de}
\affiliation{Max-Planck-Institut f\"{u}r
Polymerforschung, D-55128 Mainz, Germany}
\author{Yimin Jiang}\altaffiliation{Permanent Address:
Department of Applied Physics and Heat-Engineering,
Central South University, Changsha 410083, China}
\email{jiangym@mail.csu.edu.cn}
\author{Mario Liu}\email{mliu@uni-tuebingen.de}
\affiliation{Theoretische Physik, Universit\"{a}t
T\"{u}bingen, 72076 T\"{u}bingen, Germany}
\date{\today}

\begin{abstract}
The damping of sound waves in magnetized ferrofluids is
investigated and shown to be considerably higher than in the
non-magnetized case. This fact may be interpreted as a
field-enhanced, effective compressional viscosity -- in analogy
to the ubiquitous field-enhanced shear viscosity that is known to
be the reason for many unusual behavior of ferrofluids under
shear.
\end{abstract}
\maketitle

\section{Introduction}\label{intro}
Ferrofluids \cite{rosensweig85} are colloidal suspensions of
mono- or subdomain ferrimagnetic nano-sized particles suspended
in a carrier liquid. Under the influence of an external  magnetic
field the fluid behaves paramagnetically. Among the more
remarkable flow phenomena of ferrofluids
\cite{rosensweig85,shliomis72,blums97} are the enhanced effective
shear viscosity in a static magnetic field \cite{mctague69}, or
the viscosity decrease in response to an AC-field
\cite{morozov94,bacri95,zeuner98}. Both are due to the so-called
magneto-dissipative effect, which occurs when the experimental
time scale compares to the magnetic relaxation time. In those
situations the actual magnetization ${\bf M}$ deviates
significantly from its equilibrium value ${\bf M}^{\rm eq}$. Then
the increment ${\bf M}-{\bf M}^{\rm eq}$ feeds back to the linear
momentum balance, via the magneto-viscous stress
element~\cite{shliomis72}
\begin{equation}\label{eqpi}
\Delta \Pi_{ij}=\frac{\mu_0}{2} \varepsilon_{ijk}
[{\bf H}\times ({\bf M}-{\bf M}^{\rm eq})]_k,
\end{equation}
leading to the appearance of an enhanced shear
viscosity.

When dealing with compressible flow situations  such as
sound, one needs to go beyond the approximation of
incompressibility. If sound propagates through a
magnetized ferrofluid, density oscillations couple to the
magnetization, and one expects magneto-dissipation to
become relevant as well. Sometimes the attenuation of
sound is attributed to the elevated shear viscosity
addressed above~\cite{isler78}, but this point of view
disregards the fact that Eq~(\ref{eqpi}) only contributes
in shear flow geometries, and not in the compressional
flows characteristic of sound: Taking the divergence of
the momentum balance (to derive an equation for $\nabla
\cdot {\bf v}$, the divergence of the velocity field)
eliminates the contribution of (\ref{eqpi}), since
$\nabla_i \nabla_j \Delta \Pi_{ij}=0$. So if (\ref{eqpi})
were the only magneto-dissipative term, one must conclude
that sound in magnetized ferrofluids does not experience
any additional damping, but this is incorrect: The
recently derived ferrofluid dynamics~\cite{muller01a}
contains, in addition to the expression of
Eq.~(\ref{eqpi}),
also a new, diagonal magneto-viscous stress element,
which accounts for the additional energy loss of sound
waves if the medium is magnetized. It is natural to
interpret this fact as a magnetically enhanced
compressional viscosity, in close analogy to the
magnetically enhanced shear viscosity first observed by
McTague~\cite{mctague69}. Note that although this term
was derived in~\cite{muller01a} as a stringent result of
energy and momentum conservation, it is not contained in
the standard
ferrofluid-dynamics\cite{rosensweig85,shliomis72,blums97}.

While the well-known tensor element of~(\ref{eqpi})
is non-vanishing only if the deviation $({\bf M}-{\bf
M}^{\rm eq})$ and the field ${\bf H}$ point in
different directions, the new, diagonal stress
element---writable as ${\bf H} \cdot ({\bf M}-{\bf
M}^{\rm eq}) \, \delta_{ij}$ in the case of linear
constitutive relationship---remains non-vanishing
even if both are parallel to each other. This is
exactly the situation characteristic for the
propagation of sound. Provided the sound frequency
$\omega$ does not greatly exceed the inverse magnetic
relaxation $\tau$ of the ferrofluid, a perceptible
extra damping is predicted, several orders of
magnitude larger than estimated by previous
works~\cite{parsons75,henjes94}.

It is worth pointing out that the new diagonal stress element may
of course be disregarded in incompressible flow situations. In
these cases the pressure $p({\bf r},t)$ is determined by the
condition $\nabla \cdot {\bf v}=0$. A diagonal stress element
such as the above term then re-normalizes $p$, but leaves the
velocity profile ${\bf v}({\bf r},t)$ unchanged. Consideration of
flow configurations such as those in sound, on the other hand,
require the elimination of the incompressibility condition. This
is done by adding the continuity equation for mass and adopting a
viscous term proportional to $\nabla \cdot {\bf v}$ in the
Navier-Stokes equation. In ordinary liquids, the pressure is now
determined by the thermodynamic equilibrium relation
$p=p(\rho,T)$ as a function of density $\rho$ and temperature $T$.
But this is less simple in electrically polarized or magnetized
liquids, where the concept of pressure is rather
ill-defined~\cite{rosensweig85,henjes94,liu00}. The fact that the
energy spent (or gained) by compressing a magnetized ferrofluid
depends on the direction at which the force is applied implies
that one needs to employ the more general concept of stress, and
must especially be careful in handling the diagonal stress, as
will be outlined below.

Several recent investigations  on sound propagation
in magnetizable fluids  follow a more mesoscopic
approach, considering relative particle motions
within clusters, aggregates or chains.
Taketomi~\cite{taketomi85} attributes the anisotropy
of the sound attenuation coefficient to two types of
motions performed by the ferrous colloidal particles
in the fluid, rotational and tranlational.
Nahmad-Molinari et al. \cite{nahmad} investigated the
propagation of collective modes through a
magnetorheological slurry (with micron-sized grains)
and observed two independent modes. The slower one,
with a large amplitude, was considered to be a
compressional mode similar to what is found in porous
fluid-saturated media. Later, Brand and
Pleiner~\cite{brand01} attributed this mode to a wave
propagating along the particle chains.

Others authors~\cite{review} pursue a macroscopic, hydrodynamic
treatment similar to the approach employed here.
Parsons~\cite{parsons75} considered ferrofluids subject to strong
magnetic fields and took the vector of saturated magnetization as
similar to the director in nematic liquid crystals. As a result,
he found that magnetically induced relative corrections to the
sound velocity are small, around $10^{-5}$. As discussed by
Henjes~\cite{henjes94}, he worked with a purely mechanical
pressure ignoring electromagnetic contributions in the diagonal
stress. Using proper hydrodynamics, Henjes~\cite{henjes94} found
that the magnetically induced corrections to the sound velocity
are small, again of order $10^{-5}$. She also argued that since
typical magnetic relaxation times are of the order of $\tau
\approx 10^{-6}s$, magneto-dissipation is small for acoustic sound
frequencies up to $\omega/(2 \pi)= 20$ kHz. This is correct, but
it does not imply that magneto-dissipation can be entirely
ignored, because all dissipation mechanisms derive from fast
characteristic times, and the question is one of relative weight.
Summarizing, previous theoretical investigations on sound
propagation in ferrofluids do not account for
magneto-dissipation. The present manuscript does, and the result
is: In the hydrodynamic frequency regime, $\omega \tau \simeq 1$,
even moderate magnetic fields will induce extra damping of
approximately $10\%$.

\section{The Starting Equations}\label{system}
To quantify the damping in compressional flow situations,
propagation of sound waves through homogeneously magnetized
ferrofluids will be investigated.  To streamline the
consideration and to focus on the basic physics, we shall
implement the following simplifications: (i) Only the leading
order magnetic field effect ${\mathcal O}(H^2)$ on the
attenuation of sound is considered. This especially implies the
linear constitutive relation, ${\bf M}^{\rm eq}=\chi {\bf H}$.
Moreover, the complication \cite{muller01} that sound in
magnetized ferrofluids is generally accompanied by shear waves
need not be considered. Although this coupling gives rise to
rather surprising phenomena~\cite{muller01}, it contributes only
at ${\mathcal O}(H^4)$ to the dispersion of sound (see below).
(iii) We consider sound propagation and shear diffusion in the
adiabatic limit. Adiabaticity means $\delta {\tilde s}\equiv\delta
(s/\rho)=0$, rather than $\delta T=0$ as is the case in the
isothermal limit.  Adiabaticity is valid because in
ferrofluids, shear diffusion and sound are usually fast processes
on the time scale of heat conduction: The Prandtl number $P$,
given by the quotient of characteristic thermal diffusion time
over viscous diffusion time, or equivalently, by kinematic
viscosity over heat diffusivity, $P=\nu/\kappa$, is usually of
the order of 10-100. (Depending on the ferrofluid, we have $\nu
\approx 10^{-6}-10^{-3}m^2/s$, and $\kappa\approx
10^{-7}-10^{-5}m^2/s$.) The same argument holds for solutal
diffusion processes which are slower than shear diffusion and
sound by a factor $P/L$, where $L$ is the Lewis number. For
ferrofluids we typically have $L={\cal O}(10^{-4})$.

Below it will be discussed in more details that the magnetic
susceptibility $\chi$, usually taken as a function of $T$ and
$\rho$, must then be considered as a function of entropy per unit
mass ${\tilde s}$, in addition to $\rho$. Adiabaticity is a valid
approximation here because in ferrofluids, shear diffusion and
sound are usually fast processes on the time scale of heat
conduction. (iv) Sound waves up to the MHz-range are weakly
damped. The spatial decay length $\alpha^{-1}$ of the complex
wave number $k=\omega/c+i \alpha$ exceeds the wave number $2 \pi
c/\omega$ by many orders of magnitude \cite{footnote1}. Under
those circumstances it is the custom to account for all damping
mechanisms to linear order. (v) This paper focuses on sound
attenuation. The tiny correction to the sound velocity is
disregarded. An order of magnitude estimate for the magnetically
induced correction yields $\Delta c \simeq (\mu_0 \chi
H^2/\rho)^{1/2}$. Even at the highest magnetic field strength
considered here, one gets $\Delta c/c < 10^{-4}$.

The unperturbed state of the ferrofluid is given by a
homogeneously magnetized ferrofluid at rest, with density $\rho$
and equilibrium magnetization ${\bf M}^{\rm eq}=\chi {\bf H}$,
where $\chi$ is the magnetic susceptibility. To describe small
amplitude sound excitations we introduce deviations form this
state $\delta \rho$, ${\bf v}$, $\delta {\bf H}$, $\delta {\bf
B}$ and $\delta {\bf M}$ proportional to a plane wave with wave
vector ${\bf k}$. In particular the velocity field is taken as a
longitudinal sound mode in the form
\begin{equation}
{\bf v}\propto \frac{{\bf k}}{k} \, e^{i({\bf k}\cdot {\bf r} +
\omega t)}. \label{eq1}
\end{equation}
The equations of motion governing the ferrofluid dynamics have
recently been derived on the basis of the conservation laws and
symmetries \cite{muller01a}. The density field $\rho({\bf r},t)$
obeys as usual the continuity equation
\begin{equation}
\partial_t \rho + \nabla_j (\rho v_j)=0.
\label{eq2}
\end{equation}
The equation for the magnetization reads
\begin{equation}
\frac{{\rm d}}{{\rm d} t} M_i - \lambda_1 M_i v_{ii} - \lambda_2
M_j v_{ij}^0 + ({\bf M} \times {\bf \Omega})_i = \frac{-\chi
}{\mu_0 \tau} h_i, \label{eq3}
\end{equation}
where ${\rm d}/{\rm d}t=\partial_t + {\bf v} \cdot \nabla$ and
${\bf \Omega}=(\nabla \times {\bf v})/2$ is the
vorticity of the flow. The contributions proportional
to $\lambda_{1}$ and $\lambda_2$ appear with the
applied field breaking the isotropy of the system.
These two terms reflect the fact that -- in addition
to the vorticity ${\bf \Omega}$ -- compressional and
elongational flow, denoted respectively as
$v_{ii}=\nabla \cdot {\bf v}$ and
$v_{ij}^0=\frac{1}{2}(\nabla_i v_j + \nabla _j v_i -
\frac{2}{3} \delta_{ij} v_{kk})$, contribute to the
dynamics of ${\bf M}$. Further terms associated with
the uniaxial symmetry (see the terms proportional to
$\lambda_3$ and $\lambda_4$ in Eq.~(13) of
Ref.\cite{muller01a}) have been omitted on the left
hand side of Eq.(\ref{eq3}), as they are of higher
order in the magnetic field. The increment
\begin{equation}
- \frac{\chi}{\mu_0 \tau} {\bf h}=- \frac{\chi}{\mu_0
\tau}({\bf B}^{\rm eq}-{\bf B})=- \frac{1}{\tau}
({\bf M}- \chi{\bf H}), \label{eq4}
\end{equation}
with ${\bf B}^{\rm eq}=\mu_0 {\bf M}(1+\chi)/\chi$,
accounts for magneto-dissipative relaxation,  on the
time scale given by $\tau$.

The magnetic field variables ${\bf H},{\bf B},{\bf
M}$ are defined in SI-units as usual, with $\mu_0$
the vacuum permeability. For the evolution of the
magnetic fields we adopt the static Maxwell equations
$\nabla \cdot {\bf B}=\nabla \times {\bf H}=0$. With
the plane wave behavior similar to Eq.~(\ref{eq1})
the fluctuations are related in the following manner
\begin{equation}
\delta {\bf H}=-\delta {\bf M}_\parallel; \quad
\delta {\bf B}=\mu_0 \delta {\bf M}_\perp.
\label{eq5}
\end{equation}
Here the indices $\parallel$ and $\perp$ refer to the respective
directions relative to the propagation direction ${\bf k}$ of the
wave.

The balance equation for the linear momentum reads
$\partial_t \rho v_i + \nabla_j \Pi_{ij}=0$,
%
with the stress tensor \cite{muller01a}
\begin{eqnarray}
\label{eq7} \Pi_{ij}= [-u + sT + \mu \rho + {\bf
H}\cdot {\bf B} \\
 \nonumber - (\lambda_1 - \frac{\lambda_2}{3})
{\bf h} \cdot {\bf M}] \delta _{ij}
 -\Pi_{ij}^{vis}-H_i B_j\\
 \nonumber -\frac{ \lambda_2}{2} (M_i h_j + M_j h_i) +
\frac{1}{2} (h_i M_j - h_j M_i).
\end{eqnarray}
To avoid misunderstandings  of what is meant by the
''pressure at non-zero magnetic field strength'', the
diagonal element is written in terms of the density
of total energy $u$, the entropy density $s$, and the
chemical potential $\mu$. The viscous stresses
$\Pi_{ij}^{vis}= 2 \eta_1 \, v_{ij}^0 + \eta_2 \,
\delta_{ij} v_{kk}$ are taken as usual with the shear
viscosity $\eta_1$ and the volume viscosity $\eta_2$.
The terms proportional to $\lambda_{1,2}$ are counter
terms to those of Eq.~(\ref{eq3}), they are
constrained by the Onsager symmetry relations.

To make contact to previous formulations of the stress tensor
\cite{rosensweig85,shliomis72} we have to switch for a moment to
$T$ rather than ${\tilde s}$ as an independent variable. Then the
square bracket in Eq.~(\ref{eq7}) can be recast in terms of the
thermodynamic relation for the pressure at zero magnetic field
$p_0(\rho,T)$
\begin{equation}
p_0+\mu_0 \frac{H^2}{2} -(\lambda_1 - \frac{1}{3} \lambda_2+1)
{\bf h} \cdot {\bf M} + \frac{\mu_0 M^2}{2 \chi} (1
-\frac{\rho}{\chi}\frac{\partial \chi}{\partial \rho}),
\label{eq8}
\end{equation}
where $\chi=\chi(\rho,T)$. Note that magneto-dissipation,
proportional to  ${\bf h}\cdot {\bf M}$,  remains finite even if
the transport coefficients $\lambda_1$ and $\lambda_2$ (not yet
measured) should be negligibly small.  As outlined in
Sec.~\ref{intro}, this term arrives cogently during the
derivation of the stress tensor and accounts for
magneto-dissipative processes if ${\bf h}$ is parallel to the
equilibrium magnetization ${\bf M}^{eq}$. This is crucial for
situations where ${\bf M}$ and ${\bf H}$ oscillate co-linearly
but with a temporal phase lag. (Recall that the customary
magneto-dissipative term given by Eq.~(\ref{eqpi}) drops out if
${\bf M}$ and ${\bf H}$ are parallel to each other.)

\section{Results}\label{results}
\subsection{Dispersion of isothermal sound waves}
We now return to the adiabatic formulation, where
$\chi=\chi(\rho,{\tilde s})$. Using Eq.(\ref{eq2}) and the
longitudinal plane wave velocity field (\ref{eq1}) the
magnetization fluctuation $\delta {\bf M}=\delta {\bf
M}_\parallel + \delta {\bf M}_\perp$ is related to the density
variations by
\begin{eqnarray}\label{magnet}
\delta {\bf M}_\parallel ={\bf M}_\parallel \left \{
\frac{(\rho/\chi)\partial\chi/\partial\rho - i \omega
\tau (\lambda_1 + \frac{2 }{3}\lambda_2)}{1+\chi+ i
\omega \tau} \right \} \frac{\delta \rho}{\rho},
\label{eq10}
\end{eqnarray}
\begin{eqnarray}
\delta {\bf M}_\perp ={\bf M}_\perp \left \{
\frac{(\rho/\chi)\partial\chi/\partial\rho - i \omega
\tau (\lambda_1 - \frac{1 }{3}\lambda_2)}{1+ i \omega
\tau} \right \}  \frac{ \delta \rho}{\rho}.
\label{eq11}
\end{eqnarray}
The real and imaginary part of $\delta {\bf M}$ is
associated with magnetically induced corrections to
the sound velocity and attenuation, respectively.
Substituting Eqs.(\ref{eq10},\ref{eq11}) into the
divergence of the momentum balance yields the
following complex dispersion relation for sound waves
in magnetized ferrofluids
\begin{equation}
k=\frac{\omega}{c_s} -  i \, \frac{\omega^2}{2 \rho c_s^3} \left
( \frac{4}{3} \eta_1 + \eta_2 + \eta_m \right ),\label{eq12}
\end{equation}
where $c_s^2=\partial p_0(\rho,{\tilde s})/\partial \rho$ is the
square of the zero-field adiabatic sound velocity. Recall that --
according to approximation (v) -- magnetic corrections of $c_s$
are disregarded. The increment $\eta_m$ is given by
\begin{equation}
\eta_m= \mu_0 \tau \chi H^2 \left [
\frac{\kappa_\parallel \cos^2 \theta }{(1+\chi)^2+
(\tau \omega)^2} + \frac{\kappa_\perp \sin^2
\theta}{1+(\tau \omega)^2} \right ], \label{eq13}
\end{equation}
where
\begin{equation}
\kappa_\parallel =  \left [\frac{\rho}{\chi} \frac{\partial
\chi}{\partial \rho} +(1+\chi)(\lambda_1+\frac{2}{3}\lambda_2)
\right ]^2 , \label{eq14}
\end{equation}
\begin{equation}
\kappa_\perp=\left [\frac{\rho}{\chi} \frac{\partial
\chi}{\partial \rho} + (\lambda_1 - \frac{1 }{3}\lambda_2)\right
]^2. \label{eq15}
\end{equation}
$\eta_m$ can be interpreted as a ''magnetic extra viscosity''.
The expression for $\eta_m$ is clearly anisotropic, and $\theta$
denotes  the angle between the applied magnetic field ${\bf H}$
and the direction ${\bf k}$ of propagation. Note also that
$\eta_m$ is frequency dependent, being maximal at $\tau \omega
\to 0$ and vanishing in the high frequency limit $\tau \omega \gg
1$. Eqs.(\ref{eq14},\ref{eq15}) can be simplified if the magnetic
susceptibility is proportional to the density thus
$(\rho/\chi)\partial \chi /\partial \rho \approx 1$. For a rough
estimate, take the data \cite{embs00} for a high viscosity
hydrocarbon based ferrofluid (APG 933, Ferrofluidics): $\chi
\simeq 1.1$, $\eta_1 \simeq 0.5 $Pas, and $\tau \simeq 0.55$ms,
in addition to $\lambda_1=\lambda_2=\eta_2=0$ (for lack of better
information). Then a $100$Hz-sound wave propagating in an applied
magnetic field of say $H=10^4 {\rm A}/{\rm m}$ (this is a field
strength at which most ferrofluids still obey linear constitutive
relations) experiences a magneto-viscous extra damping of $\simeq
7\%$ at the parallel orientation ${\bf k}\parallel {\bf H}$ (i.e.
$\theta=0$) and almost $9\%$ at the transverse setup
($\theta=90^\circ$). The same estimate applies to a ferrofluid of
similar microscopic make-up, but at a viscosity of $\eta_1 \simeq
5$ mPas and  a frequency of $10$kHz. (Here we assume $\tau
\propto \eta_1$, valid for Brownian particles, i.e. when the
particle's magnetic moment is fixed to the crystallographic
orientation). Damping increments of this size should be
detectable in a careful sound wave experiment. Moreover, by
scanning the $\theta$-dependence of $\eta_m$, it should be
possible to obtain information on the transport coefficients
$\lambda_1$ and $\lambda_2$.

\subsection{Adiabatic versus isothermal susceptibility}
Knowing the dependence of the magnetic susceptibility as a
function of density and temperature, $\chi(\rho,T)$, the
derivative $\rho \frac{\partial \chi}{\partial \rho}(\rho,{\tilde
s})$ in the above equations can be expressed as follows
\begin{equation}
\rho \frac{\partial\chi(\rho,{\tilde
s})}{\partial\rho} = \rho
\frac{\partial\chi(\rho,T)}{\partial\rho} + T
\frac{\partial\chi(\rho,T)}{\partial T}\, \frac{c_T^2
\alpha_v}{C_v}, \label{eqchi}
\end{equation}
involving both the magneto-strictive and the magneto-caloric
contributions. Here  $\alpha_v=-(1/\rho)\partial
\rho(p,T)/\partial T$ denotes the thermal expansion coefficient,
$c_T$ the isothermal sound velocity, and $C_v=T
\partial {\tilde s}(T,\rho)/\partial T$ the specific heat at
constant volume, everyone of them evaluated at zero magnetic
field.
Eq.~(\ref{eqchi}) indicates that adiabatic sound
waves involve both magneto-strictive as well as
magneto-caloric contributions. Here
$\alpha_v=-(1/\rho)
\partial \rho(p,T)/\partial T$ denotes the thermal expansion
coefficient and $C_v=T \partial {\tilde
s}(T,\rho)/\partial T$ the specific heat at constant
volume, each of them evaluated at zero magnetic
field. For a typical olefine-based carrier liquid the
dimensionless factor $c_T^2 \alpha_v/C_v$ can be
estimated by $0.3$.

\subsection{Enhanced compressional viscosity} In order to
classify the viscosity increment $\eta_m$ [Eq.~(\ref{eq13})] as an
field dependent offset to either the shear viscosity $\eta_1$ or
the volume viscosity $\eta_2$ we evaluate the entropy production
in the present setup. Following Ref.~\cite{muller01a} the total
entropy production is given by
\begin{equation}\label{entropprod}
R=2 \eta_1 (v_{ij}^0)^2 + \eta_2 (\nabla \cdot {\bf v})^2 +
\frac{\chi}{\mu_0 \tau} h^2.
\end{equation}
Computing the magneto-viscous surplus [last term in
(\ref{entropprod})] up to first order in $\omega \tau$ yields
\begin{equation}\label{surplus}
\frac{\chi}{\mu_0 \tau} h^2=\mu_0 \tau \chi H^2 \left [
\frac{\kappa_\parallel \cos^2{\theta}}{(1+\chi)^2} + \kappa_\perp
\sin^2{\theta} \right ] (\nabla \cdot {\bf v})^2.
\end{equation}
The formal similarity of (\ref{surplus}) with the second term of
(\ref{entropprod}) suggests that the ''magnetic extra viscosity''
$\eta_m$ according to Eq.~(\ref{eq13}) is to be interpreted as an
enhanced compressional viscosity $\Delta \eta_2(H)$.

\subsection{Comparison with experiments}
The experimental material on sound propagation in ferrofluids is
rather scarce. The early measurements of Chung and Isler
\cite{chung78,isler78} on a water-based ferrofluid  seem to be
the only available systematic study of the velocity and
attenuation of sound (note, however, that Skumiel's
\cite{skumiel95} later investigations reveal a strong dependence
of the sound velocity on the type of the carrier liquid, i.e
whether it is aqueous or organic). The experiments of Refs.
\cite{isler78,chung78} were carried out with $2.25$ MHz
ultra-sound,  employing pulse-echo and continuous wave methods.
The experimental data cover a wide magnetic field range from $0$
up to $2500{\rm Gauss}=2\times 10^5$A/m. Within the weak field
subrange, where linear constitutive relations hold for most
ferrofluids,  $H<10^4$A/m$=125{\rm Gauss}$ say,  the damping
increment $\alpha$ was found to increase by $1.8$dB ($\simeq
20\%$) at $\theta=0$ but to decrease (anomalous sound
attenuation) by almost $3.5$dB ($\simeq 50\%$) at
$\theta=90^\circ$. Unfortunately no information was given whether
demagnetization effects due to the cylindrical probe geometry had
been taken into account here. However, we point out that the
observed anomalous $H$-dependence does not even qualitatively
comply with the present theory. Neither the observed history
dependence of the experimental data (also detected by Gotoh et
al. \cite{gotoh84}) can be explained by the present approach. The
latter peculiarities suggests that the small ultrasound
wavelengths couple to microscopic inhomogeneities associated with
particle chains or clusters in the ferrofluid suspension.
Anisotropies, field dependencies and anomalies recorded by
ultrasound experiments therefore seem to depend on mechanisms
which are rather different from those covered by the present
hydrodynamic analysis. Owing to the lack of other pertinent
experimental data let us  nevertheless try a quantitative
comparison with the measurements of Isler and Chung
\cite{isler78}. For a rough estimate we take the viscosity of
their aqueous ferrofluid by $\eta_1=10^{-3}$ Pas and the
susceptibility by $\chi \simeq 1$ (specifications are not given).
Due to the high ultra-sound drive frequency the experiment was
operated in the limit $\omega \tau \gg 1$ and thus the expected
extra damping is quite small.  Even if the magnetic relaxation
time $\tau$ is estimated to be as small as $10^{-6}$s, the
prediction of Eq.(\ref{eq13}) at $\theta=0$ is two orders of
magnitude smaller than the empiric value. We therefore conclude
that for a reliable quantitative check of the present theory,
experiments at acoustic frequencies (where $\omega \tau \simeq
1$) would be more suitable.

\section{Discussion}\label{discussion}
The present analysis deals with the attenuation of sound in
ferrofluids, which are exposed to a weak homogeneous magnetic
field in any direction relative to the propagation. This has been
accomplished by investigating the linear dispersion of a pure
longitudinal velocity excitation. Recently, it has been pointed
out \cite{muller01} that  density excitations (sound) and
transverse velocity fluctuations (shear waves) in magnetized
ferrofluids do not evolve separately as is the case at $H=0$: At
finite $H$, sound waves may produce shear excitations and vice
versa. Clearly, if shear waves accompany sound this opens a new
attenuation mechanism, which cannot be ignored. In the remainder
of this section we shall argue why this cross-coupling remains
without consequences for the present analysis:  By taking the
curl of the momentum balance one arrives at
\begin{equation}
\label{eqvortical}
 \rho \partial_t {\bf \Omega} - \eta \nabla^2
{\bf \Omega} = \frac{1}{4} \nabla \times \nabla
\times ({\bf h} \times {\bf M}).
\end{equation}
Assuming that a sound emitter produces plane density waves $\delta
\rho(t)$ within a magnetized ferrofluid, the right hand side of
Eq.~(\ref{eqvortical}) can be recast as
\begin{equation}\label{source}
\nabla \times \nabla \times ({\boldsymbol h}\times {\boldsymbol
M}) = \frac{\tau \mu_0}{\chi} \frac{\partial \chi}{\partial \rho}
\,
 \frac{{\boldsymbol M}_\perp \times {\boldsymbol
M}_\parallel}{1+\chi} \, \nabla^2
\partial_t \delta \rho \, + \, ... \, ,
\end{equation}
thus acting as a magneto-dissipative source of vorticity ${\bf
\Omega}$. If the applied magnetic field is weak we have
$\Omega={\mathcal O}(H^2)$. Via the term ${\bf \Omega} \times
{\bf M}$ in Eq.(\ref{eq2}) this sound-made vorticity induces a
third order correction in ${\bf M}$, which -- at the considered
accuracy level ${\mathcal O}(H^2)$ --  does not affect the sound
dispersion. Note however, that a proper study of sound damping in
{\em strongly} magnetized ferrofluids (as for instance undertaken
in Ref.~\cite{parsons75}) must not ignore the complication
arising from the  magneto-dissipative cross-coupling between
compressional and shear excitations.


\begin{thebibliography}{99}
\bibitem[*]{email}hwm@mpip-mainz.mpg.de, liu@itp.uni-hannover.de

\bibitem{rosensweig85}
R.E. Rosensweig, {\em Ferrohydrodynamics}, (Cambridge
University Press, Cambridge, 1985).

\bibitem{shliomis72}M.~I.~Shliomis, Sov. Phys. JETP {\bf 34}, 1291 (1972);
Usp. Fiz. Nauk {\bf 112}, 427 (1974) [Sov. Phys. Usp. {\bf 17},
153 (1974)].
%

\bibitem{blums97} E. Blums, A.
Cebers, M.M. Maiorov, {\em Magnetic Fluids}, (Walter de
Gruyter, Berlin 1997).

\bibitem{mctague69}
J.~P.~McTague, J. Chem. Phys. {\bf 51}, 133 (1969).

\bibitem{morozov94}
M.I.~Shliomis and K.~I.~Morozov, Phys. Fluids {\bf 6},2855 (1994).

\bibitem{bacri95}J.-C. Bacri, R. Perzynski, M.I. Shliomis, G.I. Burde, Phys.
Rev. Lett. {\bf 75}, 2128 (1995).

\bibitem{zeuner98}A. Zeuner, R. Richter, I.
Rehberg, Phys. Rev. {\bf E 58}, 6287 (1998).

\bibitem{isler78} W.~E.~Isler, D.~Y.~Chung, J. Appl. Phys. {\bf 49} 1812
(1978).

\bibitem{muller01a}H.~W.~M\"{u}ller and M.~Liu,  Phys. Rev. {\bf E 64},
061405 (2001).

\bibitem{parsons75} J.~D.~Parsons, J. Phys. D {\bf 8}, 1219 (1975).
\bibitem{henjes94} K.~Henjes, Phys. Rev. E {\bf 50}, 1184
(1994). See also B.I. Pirozhkov and M.I. Shliomis, Proc.
9th All-Union Acoustic Conf. (in Russian), {\bf G}, 123
(Moscow, 1977).

\bibitem{liu00}M.~Liu and K.~Stierstadt, submitted to RMP.

\bibitem{taketomi85}S.Taketomi, J. Phys. Soc. Japan {\bf 55}, 838
(1986).

\bibitem{nahmad} Y.~Nahmad-Molinari, C.~A.~Arancibia-Bulnes, and
J.~C.~Ruiz-Su\'arez, Phys. Rev. Lett. {\bf 82}, 727 (1999).

\bibitem{brand01} H.~R.~Brand, H.~Pleiner, Phys. Rev. Lett. {\bf
86}, 1385 (2001); H. Pleiner, H.~R.~Brand, J. Mag. Mag. Mat. {\bf
85}, 125 (1990).

\bibitem{review} For a critical discussion see
Ref.~\cite{henjes94}.

\bibitem{muller01}H.~W. M\"{u}ller and M. Liu,
{\it Shear-Excited Sound in Magnetic Fluid}, submitted.

\bibitem{footnote1} The damping increment in
ordinary liquids is approximately $\alpha=\eta_1 \omega^2/(2 \rho
c^3)$, where $\eta_1$ is the dynamic shear viscosity, $\rho
\simeq 1 {\rm g}/{\rm cm}^3$ the density, and $c\simeq 1400 {\rm
m}/{\rm s}$ the speed of sound. For a viscous oil with $\eta_1
\simeq 0.5$Pas one gets $\alpha/(\omega/c) \simeq \omega \times
10^{-10}$s.
%
\bibitem{embs00}J.P.~Embs, H.W. M\"{u}ller, C. Wagner,
K. Knorr, M. L\"{u}cke, Phys. Rev. E {\bf 61}, R2196 (2000).

\bibitem{chung78} D.Y. Chung, W.~E.~Eisler J. Appl. Phys. {\bf
49} 1809 (1978).

\bibitem{skumiel95}A.~Skumiel, M.~Labowski, and T.~Hornowski,
Acous. Lett. {\bf 19}, 87 (1995).

\bibitem{gotoh84}K.~Gotoh, D.~Y.~Chung, J. Phys. Soc. Jpn. {\bf
53}, 2521 (1984).









\end{thebibliography}
\end{document}